**Coupling Remote States through the Continuum: Multi-State Fano Resonances**

Y. Yoon[1], M.-G. Kang[1], T. Morimoto[2], M. Kida[3], N. Aoki[3], J. L. Reno[4], Y. Ochiai[3], L. Mourokh[5], and J. P. Bird[1,3]

1: Department of Electrical Engineering, University at Buffalo, the State University of New York, Buffalo, NY 14260-1920, USA

2: Advanced Device Laboratory, RIKEN, 2-1 Hirosawa, Wako, Saitama 351-0198, Japan

3: Graduate School of Advanced Integration Science, Chiba University, 1-33 Yayoi-cho, Inage-ku, Chiba 263-8522, Japan

4: CINT Science Department, Sandia National Laboratories, P.O. Box 5800, Albuquerque, NM 87185-1303

5: Department of Physics, Queens College of CUNY, 65-30 Kissena Blvd., Flushing, NY 11367, USA

We demonstrate a fully-tunable multi-state Fano system in which remotely-implemented quantum states interfere with each other through their coupling to a mutual continuum. On tuning these resonances near coincidence a robust avoided crossing is observed, with a distinctive character that confirms the continuum as the source of the coupling. While the continuum often serves as a source of decoherence, our work therefore shows how its presence can instead also be essential to mediate the interaction of quantum states, a result that could allow new approaches to engineer the collective states of nanostructures.

While resonances abound in classical physics, quantum-mechanical wavefunction superposition allows for new manifestations of resonant behavior, most notably in the case of the Fano resonance. Originally revealed in atomic autoionization [1-9], but inherent also to solid-state systems [10-22], this resonance arises from the distinctive coupling of a continuum to a discrete quantum level, thereby providing two interfering pathways for a transition between some initial and final state. Although the details of this resonance have been studied for more than half a century, Fano also predicted [1] that the continuum-based coupling underlying it should allow for more complicated behavior, namely "multi-state" Fano resonances involving the interaction of several quantum states. Examples of such behavior are rare, however, with the phenomenon of $q$-reversal due to intruder-states in Rydberg atoms being most notable [6-9]. In this effect, the discrete Fano resonances arising from a specific manifold of levels in a Rydberg atom are strongly modified by the existence of other manifolds in the same energy range. In spite of its complicated characteristics, the $q$-reversal is a single-atom effect, in which the levels involved in the multi-state resonance arise within the same atom. In this Letter, however, we demonstrate a very different form of multi-state Fano resonance, in which two distinct states are realized on spatially-remote nanostructures, and interfere with each other through an interconnecting, and also separately implemented, continuum. While each state exhibits a Fano resonance due to its own interaction with the continuum, tuning the two states near coincidence allows an avoided crossing of their resonances to be observed, indicating the formation of an extended molecular state that is "bonded" through the continuum. This mesoscopic effect is found to be surprisingly robust, particularly if we consider that the quantum states involved are coupled indirectly through the continuum, rather than being directly connected. While the continuum often serves as a source of decoherence, our work therefore shows how its presence can instead also be essential to mediate the interaction of quantum states, a result that could allow new approaches to engineer the collective states of nanostructures.

To realize a multi-state Fano system, we take advantage of our recent demonstrations of a flexible approach to the implementation of Fano resonances in solid-state nanostructures. Specifically, we use quantum point contacts (QPCs) that are realized by applying a depleting voltage to "split" Schottky gates, thereby forming a nanoconstriction in a high-mobility two-dimensional electron gas (2DEG). When this constriction is about to pinch-off, the strongly reduced electron density inside it is expected to enhance many-body interactions, leading to the formation of a quantum-dot-like bound state (BS) in its self-consistent potential [23,24]. We have shown that a BS in one QPC (referred to here as the "swept" QPC) can induce a Fano resonance in the conductance ($G_d$) of a second QPC (referred to as the "detector"), when the two are coupled to each other through a common 2DEG [25-29]. The Fano resonance arises when the gate voltage ($V_s$) applied to the swept QPC drives its BS through the 2DEG/detector Fermi level, creating a correlation [26] in which electrons tunnel repeatedly between the BS and the detector. Consistent with this mechanism, we have furthermore shown that the $q$-parameter [1] quantifying the BS-continuum coupling, and reflected in the asymmetry of the Fano resonance, can be systematically controlled in experiment, by varying the path length through the 2DEG between the BS and the detector [28].

In this work, we use QPCs as an "on-demand" source of quantum states to implement the configurations of interacting BSs shown in Figs. 1(b) and 1(c). In both configurations, two BSs are formed on remote QPCs ($BS_1$ in the "swept" QPC and $BS_2$ in the "control" QPC), separated from each other by a distance of about 300 nm, and connected through an intervening region of 2DEG. The device used for these studies was that of Refs. [27-29], and was implemented in a GaAs/AlGaAs quantum well (Sandia sample EA750). The 2DEG was of density $2.3\times10^{11}$ cm$^{-2}$, mobility $4\times10^6$ cm$^2$/Vs, mean free path 31 microns, and Fermi wavelength 53 nm (all at 4.2 K). The mean free path decreased to 4 microns by 77 K, but this was still much longer than even the largest inter-QPC separation in the device. The device had eight contacts positioned around the perimeter of its Hall bar, which allowed us to independently determine the conductance of the various QPCs. These measurements were performed at 4.2 – 40 K, using lock-in detection (11 Hz) and a fixed excitation of 30 µV. In each configuration, unused gates were held at ground potential and our prior work [27,28] suggests these gates exert minimal influence on the resulting device behavior.

Figures 1(d) and 1(e) show contours generated from experiments in which a fixed voltage was applied to the gates of the detector, after which its conductance ($G_d$) was measured while varying the gate voltages $V_s$ and $V_c$. The limits of these contours were chosen to ensure that they span the parameter space for which BS formation is expected to occur in both the swept and control QPCs. Each contour shows two Fano resonances, labeled $R_1$ and $R_2$ to indicate their connection to the presence of a BS in the swept and control QPCs, respectively. Consistent with our analysis of the single-BS/detector interaction [28], the lineshape of the resonances is only slightly asymmetric (see, for example, the boundaries of the contours). This indicates that the Fano coupling of the BSs to the detector is weak (i.e. $|q| > 1$ [1]), which can be attributed to the relatively large separation between them [28]. Our assignment of the resonances to BSs in specific QPCs was made by comparing the positions of $R_1$ and $R_2$ to the results of separate measurements of the conductance of the swept and control QPCs themselves. In this way, we found $R_1$ and $R_2$ to be strongly correlated to the pinch-off of the swept and control QPCs, respectively, indicating their association with $BS_1$ and $BS_2$. These assignments are actually consistent with the different dispersions that $R_1$ and $R_2$ exhibit in Figs. 1(d) and 1(e). Since $BS_1$ is defined by the gate voltage $V_s$ alone, its resonance ($R_1$) occurs at an almost constant value of $V_s$ when $V_c$ is varied. $BS_2$, on the other hand, is defined by both $V_s$ and $V_c$, so that making $V_c$ more negative shifts the pinch-off of the control QPC, and so $R_2$, to less-negative $V_s$.

The continuum mediated coupling in our system gives rise to an unusual behavior of the resonances in Figs. 1(d) and 1(e). Based on the manner in which $R_1$ and $R_2$ evolve with gate voltage, it appears that they should exhibit a crossing for a particular combination of $V_s$ and $V_c$ (shown circled). Both contours exhibit a clear anti-crossing, however, whose form is unusual since it has a missing branch (see the dotted lines). The observation of this "three-branched" anti-crossing is a critical result, which demonstrates that the coupling of the two BSs in our system arises primarily through the interconnecting 2DEG. In both Figs. 1(d) and 1(e), the missing branch corresponds to a range of parameter space where one is trying to detect the resonance due to the distant BS ($BS_1$), while the intervening (control) QPC that separates it from the detector is fully pinched-off (as indicated in the lower inset to Fig. 2(c)).

To "recover" the missing branch of the anti-crossing, in spite of the quenching of the continuum-based coupling between the detector and the distant BS, it is necessary to change the measurement configuration to use the pinched-off control QPC as a detector. This is demonstrated in Fig. 2(b), in which, in spite of its extremely low conductance ($G_c$), the control QPC shows a small [27] peak that is correlated very closely to the pinch-off of the swept QPC (Fig. 2(a)) (the step-like change that accompanies the conductance peak in this figure only becomes apparent when $G_d$ is close to pinch-off, and is thought to arise from a change in electrostatic screening that occurs when the swept QPC fully pinches-off) This peak can therefore be used to provide detection of $BS_1$, and in Fig. 2(c) we combine the results of such measurements with those of Fig. 1(d). The two different data sets appear consistent with each other, and allow the full structure of the anti-crossing to become clear. Based on the analyses of Figs. 1 and 2, we therefore conclude that we have successfully implemented a multi-state Fano system whose separate quantum states are effectively "bonded" through a continuum.

Further evidence for interaction of the Fano resonances is provided by their temperature ($T$) dependence. Fig. 3(a) shows measurements of the multi-state Fano system that were performed by adjusting $V_c$ to bring $R_1$ and $R_2$ into close proximity at low temperatures (these measurements were performed during a thermal cycle more than a year after those of Fig. 1, demonstrating the excellent reproducibility of our experiment). In Fig. 3(b), we plot the $V_s$-position of the two resonances as a function of $T$. Above 10 K, the resonances shift to more-negative $V_s$ with increasing $T$, a trend discussed previously for the single-BS/detector configuration [27]. When the temperature is lowered below 10 K, however, we clearly observe the onset of peak repulsion, and this is also demonstrated in the inset to Fig. 3(b) which plots the separation of the two peaks ($\Delta V_{pp}$) as a function of $T$. The repulsion is only found with $V_c$ and $V_s$ adjusted so that the two resonances are in close proximity, as we confirm in the upper inset to Fig. 2(c). Here we show the temperature-dependent variation of the resonance positions measured with them detuned significantly from the anti-crossing, and find no evidence for peak repulsion.

The phenomenology of Fano resonances has been widely investigated in atomic and mesoscopic systems, with numerous studies having addressed the interference of a single discrete state (or even a series of isolated states) with a continuum [22]. More complicated behavior arises if resonances overlap, however, as has been discussed for intruder states in Rydberg atoms [6,7], and the Coulomb-modified Fano resonances of quantum dots [15]. While in both of these examples the overlapping resonances originate within the same atom or dot, the interacting quantum states in this study are implemented separately, and interact with each other through an additionally-separate continuum. Among the mechanisms that could arise from such coupling include: level hybridization due to the mutual overlap of the BS wavefunctions; spin-spin exchange between electrons populating the BSs, or; a Ruderman-Kittel-Kasuya-Yosida (RKKY) interaction of these electrons. The distinctive survival of the peak repulsion to temperatures as high as 10 K (Fig. 3) indicates an interaction energy much larger than that typical of these mechanisms, however, especially if we consider that the QPCs are separated through an intervening 2DEG. While the detailed nature of the coupling is not clear at present, our understanding of the Fano character of the resonance that arises from the interaction of the detector with just a single BS [25-29] suggests the mechanism that is illustrated schematically in Fig. 1(a). In this one-electron effect, supported by the extremely high-mobility of the 2DEG, electron partial waves emerging from the detector scatter from the two BSs, giving rise to separate Fano resonances in the detector conductance. According to such a mechanism, the avoided crossing of the resonances arises from the strong spatial overlap of the two Fano paths. That is, the continuum not only interferes with the BSs to yield separate Fano resonances, but also mediates their interaction, inducing the formation of an extended artificial molecule [16,30].

In conclusion, we have provided a demonstration of a fully-tunable multi-state Fano system in which remotely-implemented quantum states interfere with each other through their coupling to a mutual continuum. On tuning these resonances near coincidence a robust avoided crossing is observed, with a distinctive "three-branched" character that confirms the continuum as the source of the coupling. While the continuum often serves as a source of decoherence, our work therefore shows how its presence can instead also be essential to mediate the interaction of quantum states, a result that could allow new approaches to engineer the collective states of nanostructures.

This work was supported by the Department of Energy (DE-FG03-01ER45920) and was performed, in part, at the Center for Integrated Nanotechnologies, a U.S. DOE, Office of Basic Energy Sciences nanoscale science research center. Sandia National Laboratories is a multi-program laboratory operated by Sandia Corporation, a Lockheed-Martin Company, for the U. S. Department of Energy under Contract No. DE-AC04-94AL85000.

**FIGURE CAPTIONS**

**Fig. 1** (a) Schematic illustration of the key components of the Fano system and their interactions. (b), (c) Scanning electron micrographs showing how the device is configured in different experiments to realize the Fano system. Light regions are the Ti/Au gates that can be used to implement QPCs in different spatial arrangements. The dark areas are the surface of the GaAs/AlGaAs heterostructure. (d) Measured variation of $G_d$ at 4.2 K, using the combination in (b), as a function of the gate voltages $V_c$ and $V_s$. (e) As in (d), except for the combination in (c). Color variation from blue to red corresponds to a conductance change of $0 – 0.33 \times 2e^2/h$ in (d), and $0 – 0.21 \times 2e^2/h$ in (e). In both (d) and (e), a slowly-varying background [28] with an average value of approximately $3.5 \times 2e^2/h$, has been subtracted from $G_d$.

**Fig. 2** (a) $G_s(V_s, V_c)$. (b) $G_c(V_s, V_c)$. White dotted lines in (a) and (b) are guides to the eye that follow the same variation. Black dotted line at front of contour (b) is a guide to clarify the lineshape of $G_c(V_s)$. (c) Evolution of resonances $R_1$ and $R_2$ as a function of $V_c$ and $V_s$. Open symbols are data determined with control QPC as detector. Dotted lines are guides to eye. Measurements are for the same gate configuration as in Fig. 1(b). The upper inset to Fig. 2(c) shows the measured variation of the peak position ($V_s$) of $R_1$ and $R_2$, obtained in the same experiment as that shown in Fig. 3, except that now $V_s$ and $V_c$ are configured such that the two resonances are far from their avoided crossing. The lower inset indicates the configuration of the QPCs when the missing branch of the ant-crossing is observed (relative to the configuration of Fig. 1(b)).

**Fig. 3** Measurements of the double-peak structure at different temperatures. The measurement configuration is the same as that of Fig. 1(b), and a slowly-varying background [28], with an average value of approximately $3.5 \times 2e^2/h$, has been subtracted from $G_d$. $V_c$ was held fixed at -1.214 V in these measurements, which were performed on a different thermal cycle to those of Figs. 1 and 2. (b) $V_s$ position of the two peaks as a function of temperature. Symbols in the figure correspond to the peak assignment indicated in Fig. 3(a). The inset plots the separation of the two peaks as a function of temperature.

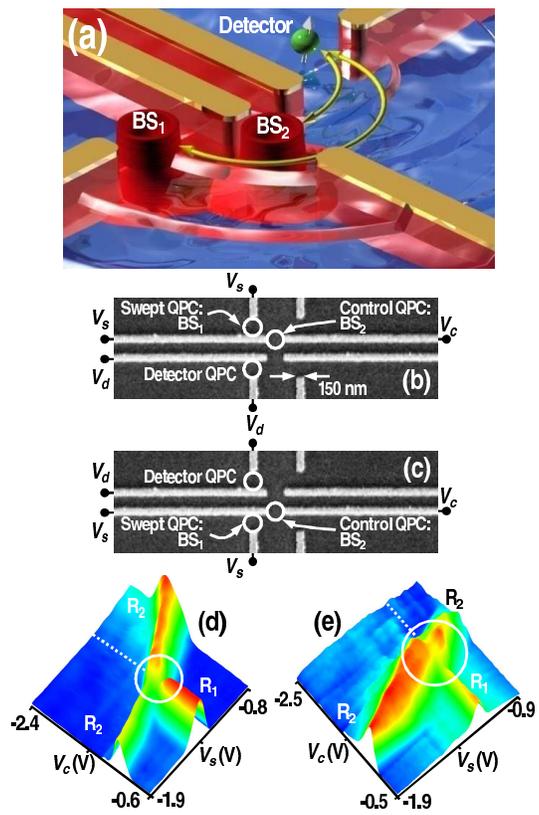

Figure 1    LL11810    02NOV2009

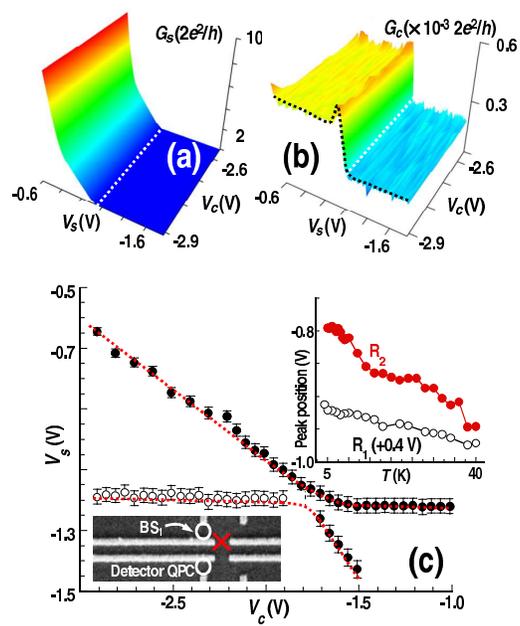

Figure 2    LL11810    02NOV2009

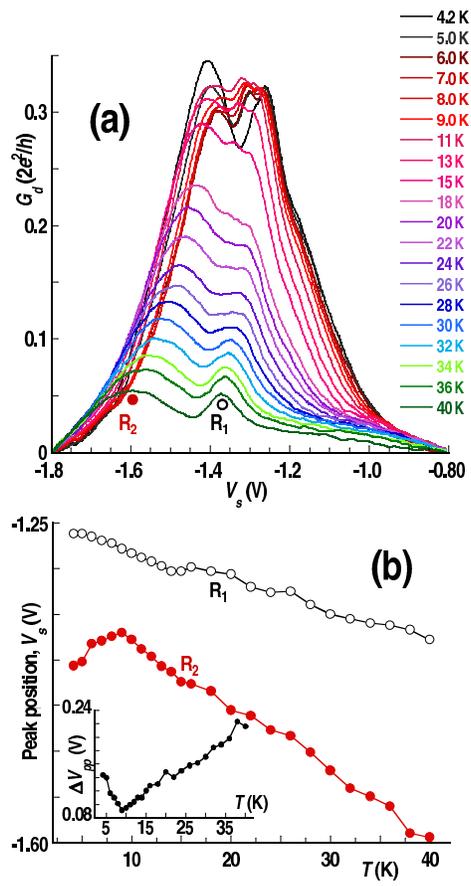

Figure 3    LL11810    02NOV2009